\documentstyle[aps,pra,floats,epsf,amssymb]{revtex}

\newcommand{\bA}{{\bf A}}
\newcommand{\bB}{{\bf B}}
\newcommand{\bH}{{\bf H}}
\newcommand{\bQ}{{\bf Q}}
\newcommand{\bL}{{\bf L}}
\newcommand{\br}{{\bf r}}
\newcommand{\bk}{{\bf k}}
\newcommand{\bq}{{\bf q}}
\newcommand{\bv}{{\bf v}}
\newcommand{\bm}{{\bf m}}

\newcommand{\ba}{{\bf a}}
\newcommand{\bb}{{\bf b}}

\newcommand{\be}{{\bf e}}

\newcommand{\cD}{{\cal D}}

\newcommand{\kB}{k_{\rm B}}
\newcommand{\half}{\frac{1}{2}}
\newcommand{\f}{\frac}

\newcommand{\Eq}[1]{Eq.~(\ref{#1})}
\newcommand{\Fig}[1]{Fig.~\ref{#1}}
\newcommand{\Ref}[1]{(\ref{#1})}

\begin{document}

\title{Superconducting Coherence and the Helicity Modulus in Vortex
Line Models}

\author{Jack Lidmar and Mats Wallin}

\address{Department of Theoretical Physics, Royal Institute of
Technology, SE-100 44 Stockholm, Sweden}

\date{\today}


\wideabs{

\maketitle
\draft

\begin{abstract}
We show how commonly used models for vortex lines in three dimensional
superconductors can be modified to include ${\bf k=0}$ excitations.
We construct a formula for the ${\bf k=0}$ helicity modulus in terms
of fluctuations in the projected area of vortex loops.  This gives a
convenient criterion for the presence of superconducting coherence.
We also present Monte Carlo simulations of a continuum vortex line
model for the melting of the Abrikosov vortex lattice in pure YBCO.
\end{abstract}


}

Phase transitions involving vortices in high temperature
superconductors are the subject of intense study both experimentally
and theoretically~\cite{Blatter}.  The enhanced thermal fluctuations
strongly alter large parts of the mean field phase
diagram~\cite{Nelson,FFH}, with new phases appearing, e.g., vortex
line liquids, vortex glass phases, etc.  A convenient quantity to
study theoretically is the helicity modulus $\Upsilon$ (or spin-wave
stiffness), which measures the free-energy increment associated with
an externally imposed twist in the phase of the superconducting order
parameter~\cite{M.Fisher}, and is proportional to the macroscopic
superfluid density.  Much work has been based on $XY$-like models,
defined in terms of this phase, where vortices appear only as
topological defects.  However, a formulation directly in terms of
vortex degrees of freedom has several advantages.  In this paper we
show how the uniform helicity modulus can be defined directly in terms
of the vortex lines, without reference to the phase.  We also show
Monte Carlo results for $\Upsilon$ and other quantities in a continuum
model of interacting vortex lines.

One of the advantages of the vortex representation is the possibility
to define the model on a continuum, and so avoid artificial pinning to
a discretization lattice.  Furthermore, one may include interactions
coupling directly to the vortex lines, such as core energies and
various types of disorder.  The vortex representation therefore allows
new parameter regimes to be reached compared to the phase
representation.  Both representations are frequently used in computer
simulations~\cite{Li-Teitel,Chen-Teitel,Simulations,Nordborg-Blatter}.
$\Upsilon$ is straightforward to calculate in the phase
representation~\cite{Li-Teitel}.  However, in the vortex
representation usually only ${\bf k \ne 0}$ fluctuations are included,
making the uniform response $\Upsilon$ exactly zero, and
extrapolations from finite ${\bf k}$ become
necessary~\cite{Chen-Teitel}.  Furthermore, when screening from gauge
field fluctuations is taken into account, $\Upsilon(\bk)\sim k^2$ for
small, but finite $\bk$, since an imposed phase twist can be
compensated by the gauge field.  Extrapolation to ${\bf k=0}$ then
gives zero, but the small-$\bk$ behavior of $\Upsilon(\bk)$ may still
be related to the Meissner effect in a superconductor and used to
detect phase transitions~\cite{Chen-Teitel}.  The methods mentioned
above involve extrapolations from the smallest available wave vectors
to ${\bf k = 0}$, thereby severely complicating the data analysis.  An
alternative may be to study winding number fluctuations, which are
related to the magnetic permeability $\mu$, but they suffer from being
difficult to equilibrate for large system sizes.

In this paper we take a different route by modifying the vortex model
in order to incorporate fluctuations with zero wave vector.  The form
of this modification is obtained using a duality transformation
between the phase and vortex representations, paying due attention to
the role of the boundary conditions.  We show that periodic boundary
conditions for the phases enter as an additional term in the
Hamiltonian of the vortex representation.  This allows direct
evaluation of the ${\bf k = 0}$ helicity modulus in terms of
fluctuations of the total net area of vortex loops, which can indeed
be finite also in the presence of screening.  The role of boundary
conditions in the duality transformation has previously been explored
in 2D lattice models~\cite{Peter,Vallat,Ney-Nifle} and 3D gauge glass
models~\cite{Bokil-Young}.  Here we generalize this idea to continuous
3D systems, with finite magnetic field, penetration depth, and
temperature.  Furthermore, we report on Monte Carlo simulations of a
continuum London model.  In contrast to previous continuum
simulations, which used 2D Bose models with planar
interaction~\cite{Nordborg-Blatter}, we take into account the full 3D
long range interaction.  Our model has the essential features to
describe the vortex lattice melting transition in pure YBCO, where a
continuum description should apply.

The starting point for our discussion is the Ginzburg-Landau theory in
the London limit, where amplitude fluctuations of the superconducting
order parameter $\Psi=|\Psi|\exp(i\theta)$ are neglected.  For
simplicity we will use an isotropic continuum description, since our
results are independent of microscopic details.  The generalization to
other cases is straightforward.  The Hamiltonian reads
\begin{equation}				\label{phase}
  H = \int_\Omega d^3\br\left\{\f{J}{2}\left(\nabla\theta-{2\pi \over
  \Phi_0} \bA \right)^2 +{\bB^2 \over 8\pi}-{\bB\cdot\bH \over 4\pi}
  \right\},
\end{equation}
where $\Omega=L_xL_yL_z$ is the size of the system, $\theta(\br)$ is
the phase of the superconducting order parameter, $\bB = \nabla \times
\bA$ is the magnetic flux density, $\bH$ is an externally applied
magnetic field, $\Phi_0=hc/(2e)$ is the flux quantum, and
$J=\Phi_0^2/(16\pi^3\lambda_0^2)$ is a coupling constant with
$\lambda_0$ the bare magnetic penetration depth.  The first term in
\Eq{phase} is the kinetic energy with the superfluid velocity $\bv =
\nabla\theta- (2\pi/\Phi_0)\bA$, while the second describes the
magnetic energy.  The partition function is obtained by integrating
over the phases $\theta(\br)$, and gauge field $\bA(\br)$, subject to
some gauge fixing condition: $Z=\int\cD\theta\cD'\bA e^{-\beta H}$.
In order to get a finite result a short distance regularization has to
be imposed, e.g., by defining the model on a lattice.  Physically this
cutoff is of the order of the Ginzburg-Landau coherence length
$\xi_0$, and gives the size of the vortex cores.

We now discuss the transformation of the model, \Eq{phase}, to a
system of interacting vortex lines in some detail.  For simplicity we
start by considering the case without any externally applied magnetic
field, $\bH=0$.  The interaction can be linearized by an integration
over an auxiliary field $\bb(\br)$, upon which the kinetic energy
becomes $\int_\Omega d^3\br J\left( i \bb\cdot\bv + \half \bb^2
\right)$.  The superfluid velocity splits into a longitudinal part,
describing the smooth spin-wave fluctuations of the order parameter,
and a transverse part describing the singular vortices: ${\bf v =
v_\parallel + v_\perp}$.  Integrating over the longitudinal part leads
to the constraint $\nabla\cdot\bb=0$, which can be enforced by setting
$\bb=\nabla\times\ba$.  After a partial integration of the first term
and subsequent integration over $\bB$ the Hamiltonian becomes
\begin{equation}				\label{2+1}
  H = \int_\Omega d^3\br J\left\{2\pi i \ba \cdot \bm + \half \left(
  \nabla\times \ba \right)^2 + \f{1}{2\lambda_0^2} \ba_\perp^2
  \right\},
\end{equation}
where $\ba_\perp$ is the transverse part of $\ba$ and $\bm(\br)$
denotes the vorticity, $\nabla \times \nabla\theta = 2\pi\bm$.
Integrating out the auxiliary field $\ba$ leaves us with the vortex
Hamiltonian~\cite{Dasgupta-Halperin,Chen-Teitel,Blatter}
\begin{equation}				\label{vortex_ham}
  H = \int_\Omega \f{K}{2} \bm(\br) \cdot V({\bf r - r'}) \bm(\br')
  d^3\br d^3\br',
\end{equation}
where $K=(2\pi)^2J$ and $V$ is the London interaction
\begin{equation}				\label{London}
  V(\br) = \f{1}{\Omega} \sum_\bk \f{e^{i\bk\cdot\br}}{\bk^2 + \lambda_0^{-2}},
\end{equation}
$k_\mu=2\pi n_\mu/L_\mu$, $n_\mu \in {\Bbb Z}$, $\mu=x,y,z$.  With
periodic boundary conditions for the phases $\theta$ we also get the
constraint of zero net vorticity $\int_\Omega\bm(\br)d^3\br=0$.

In going from \Eq{phase} to \Eq{2+1} we implicitly assumed that $\bv$
had no ${\bf k = 0}$ component, allowing us to throw away a surface
integral.  However, if the phases obey periodic boundary conditions,
$\theta({\bf r + L}_\mu) = \theta(\br)$ (where $\bL_\mu=L_\mu
\be_\mu$), there will be an additional energy term, coming from
uniform fluctuations of $\bv$.  An important point here is that the
integration over $\bA$ should not include the uniform part $\bA_0$,
since such fluctuations correspond to fluctuations in the boundary
conditions.  The additional energy is simply $H'={J\over 2\Omega}
\bv_0^2$, with $\bv_0 = \int_\Omega \bv(\br) d^3\br$.  This is now
related to the vortices as follows.  The contribution to the vorticity
$\bm$ from a single vortex loop can be written
\begin{equation}				\label{derivation}
  \bm(\br) = \oint_\Gamma \delta({\bf r - r'})d\br' =
  \nabla\times\int_S\delta({\bf r - r'})d{\bf S'},
\end{equation}
where $\Gamma$ is a contour describing the vortex loop, and $S$
denotes an oriented surface which has $\Gamma$ as a boundary.
Summing over all vortex loops gives the total vortex density.  Now,
since the vorticity is the rotation of the superfluid velocity, we may
define
\begin{equation}				\label{Q}
  \bQ={1\over 2\pi} \int_\Omega \bv d^3\br = \sum_i\int_{S_i} d{\bf
  S}_i,
\end{equation}
where the sum is over all vortices.  This has the interpretation of
the total projected net area of the vortex loops in each direction.
Due to the periodic boundary conditions, the value of $\bQ$ is
uniquely determined by the positions of the vortices only up to an
integer multiple of $\Omega/L_\mu$, reflecting the need to specify the
total phase twist of the system.  {\em Thus, the variable $\bQ$ keeps
track of the total phase twist of the system and must be independently
specified in addition to the vortices in order to completely specify
the state of the system}.  The additional ${\bf k=0}$ component of the
energy is now given by
\begin{equation}				\label{area}
  H'=K\f{\bQ^2}{2\Omega},
\end{equation}
and the total Hamiltonian is given by the sum of Eqs.~\Ref{vortex_ham}
and \Ref{area}: $H_{\rm tot}=H+H'$.

{\em External magnetic field.}--- Assume now that $n$ flux quanta
$\Phi_0$ penetrates the system in the $z$-direction.  In this case the
periodic boundary conditions for the phases have to be changed so that
the system can accommodate a net number of vortices.  For the vortices
penetrating the whole system the area $\bQ$ should now be measured
with respect to a given reference line at some arbitrary but fixed
position determined by the boundary conditions.  A possible choice is
to let $\theta(\br + \bL_\mu) - \theta(\br) = n\pi - {2\pi\over\Phi_0}
\int_\br^{\bf r + L_\mu} \bA\cdot d\br'$, with the integral taken
along a straight line across the system.  In the gauge
$\nabla\cdot\bA=0$ we may write $\bA(\br)=\bA_{\rm per}(\br) +
\half\bar\bB\times(\br-\br_0)$, where $\bA_{\rm per}$ satisfy periodic
boundary conditions and $\bar\bB=n\Phi_0\bL_z/\Omega$ is the uniform
part of the flux density.  In this case the fixed reference line goes
through $\br_0$ in the direction of $\bar\bB$.

{\em Helicity modulus.}--- The full importance of the new term in the
energy becomes evident when one considers the superfluid response of
the system.  Replacing the boundary conditions by twisted, $\theta(\br
+ \bL_\mu) \to \theta(\br+\bL_\mu) + \Theta$, leads to the replacement
$Q_\mu \to Q_\mu - \tilde{A}_{\mu}$ with $\tilde{A}_{\mu} =
\Omega\Theta/L_\mu$ in \Eq{area}.  This allows us to define the zero
wave vector helicity modulus by
\begin{equation} \label{helicity}
   \Upsilon_\mu = \f{\Omega}{K}\f{\partial^2F}
   {\partial\tilde{A}_{\mu}^2} = 1-{K\over \Omega T}
   \left<\!\left<Q_\mu^2\right>\!\right>,
\end{equation}
where $\langle\!\langle Q^2\rangle\!\rangle = \langle Q^2\rangle -
\langle Q\rangle^2$, and $F=-T\ln Z$ is the free energy.  $\Upsilon$
is non-zero in the superconducting state and vanishes at the phase
transition.  In the critical region of a continuous phase transition
it obeys the Josephson scaling relation $\Upsilon \sim \xi^{-1}$,
where $\xi(T)$ is the correlation length~\cite{M.Fisher}.

\begin{figure}[t]
\epsfxsize=1 \linewidth \epsfbox{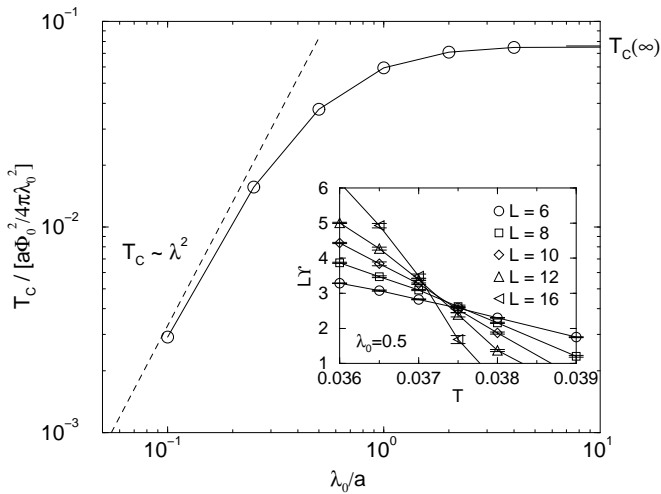} \narrowtext
\caption{
Monte Carlo results for the ($\lambda_0,T$) phase diagram for $\bB=0$.
The dotted line indicates the value of $T_c$ for the inverted
$XY$-model obtained in the limit $\lambda_0 \to 0$, $T_c \approx
Ka(\lambda_0/a)^2/3.0$ ($a$ is the lattice constant $\sim \xi_0$).  In
the opposite limit, $\lambda_0\to\infty$, $T_c \approx
3.0Ka/(2\pi)^2$.  Inset: $T_c$ is located at the intersection of
curves for $L\Upsilon$ for different $L$.
}
\label{fig.phase}
\end{figure}

\begin{figure}[t]
\epsfxsize=1 \linewidth \epsfbox{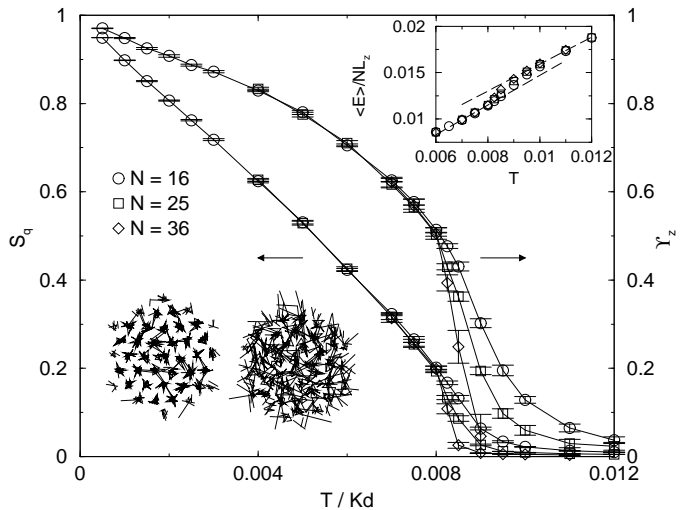} \narrowtext
\caption{
Helicity modulus $\Upsilon_z$ and structure function $S_q$ at an
ordering vector of the vortex lattice.  Inset shows the jump in energy
per vortex and layer at the transition.  Also shown are two typical
snapshots from the simulation, one below, and one above $T_c \approx
0.008Kd$.
}
\label{fig.2}
\end{figure}

The ordinary vortex Hamiltonian, \Eq{vortex_ham}, without the
additional area term $H'$ is recovered by integrating over $\Theta$
and thus corresponds to having fluctuating boundary
conditions~\cite{Peter}.  In this case the ${\bf k=0}$ response is
zero.  Fluctuations in the winding number ${\bf W}=\int_\Omega
\bm(\br) d^3\br$, are obtained only if the ${\bf k=0}$ component of
the magnetic flux density $\bB$ is allowed to fluctuate.  Then the
magnetic permeability $\mu_\nu = 4\pi\partial \left<B_\nu\right>
/\partial H_\nu = \left<\!\left< B_\nu^2\right>\!\right>/\Omega T =
\left<\!\left< W_\nu^2 \right>\!\right>\Phi_0^2/\Omega T$ can be
calculated.

{\em Monte Carlo simulations.---} 
To clearly demonstrate the practical usefulness of these ideas we now
discuss Monte Carlo (MC) simulations.  The inclusion of the area term
in a simulation is straightforward.  With the model defined in terms
of vortices, the Monte Carlo moves consist of deformations of the
vortex lines and (possibly) creations and destructions of closed
loops.  The change in the projected area coming from these local
updates are accumulated in $\bQ$, and the change in total energy,
including the area term \Eq{area}, must be used to calculate the
transition probabilities of the Markov chain.  Optionally, these moves
may be supplemented by global moves where $Q_\mu$ is changed by $\pm
\Omega/L_\mu$, corresponding to dragging a whole vortex across the
entire system.  The acceptance ratio for such moves can be expected to
be quite low because of the high energies involved.  Alternatively
these global moves can be integrated out exactly, leading to the
replacement of $H'$ by a periodic Gaussian: $\exp(-\beta H') \to
\sum_{M_\mu} \exp(- {1 \over 2\Omega T}{\sum_\mu
K(Q_\mu-M_\mu\Omega/L_\mu)^2})$ \cite{foot}.  Since this form of the
area term leads to a somewhat more complicated expression for the
helicity modulus it will not be used in what follows.

We present simulation results for two different models: (i) a lattice
superconductor in zero magnetic field with different values of
$\lambda_0$, (ii) a continuum vortex model of the melting of the
Abrikosov vortex lattice.  In both cases $10^5-10^6$ Monte Carlo
sweeps were used, with the initial $\sim 10\%$ discarded for
equilibration.

In \Fig{fig.phase} we present results from the first case, in which
the phase transition is continuous.  The critical temperature $T_c$ is
determined from finite size scaling of the helicity modulus
$\Upsilon(T,L)=L^{-1}\tilde\Upsilon([T-T_c]L^{1/\nu})$, as shown in
the inset.  Due to the new length scale given by $\lambda_0$, scaling
works only for rather large system sizes and corrections are clearly
visible for small sizes, but the determination of $T_c$ is still quite
accurate.  In the limits $\lambda_0 \to \infty$ and $\lambda_0 \to 0$
we recover known results, showing that our method works properly.
Furthermore, a scaling collapse of MC data for $\lambda = 0.25$ is
obtained using the expected 3D XY value $\nu \approx 2/3$.

In our continuum simulations in an applied magnetic field, we
discretize the vortex lines only along the $z$-direction, using
straight segments to interpolate between the $xy$-planes where the
positions are continuous.  We exclude overhangs and isolated loops,
which should be of importance only close to the zero field $T_c$.  In
addition to local MC moves of the positions in the $xy$-plane, we
include moves where two flux lines are cut off and reconnected to each
other, allowing different permutations of the boundary conditions to
be sampled.  We use a full 3D long range interaction, given by
\Eq{London} (with $\lambda_0 = \infty$), supplemented by a Gaussian
short distance cutoff $e^{-k^2\xi_0^2}$, which acts between the
midpoints of the vortex line segments.  The vortex lattice constant
was set to $4\xi_0$ and the layer separation $d$ to $2\xi_0$.  The
number of layers for a system of $N$ vortices was set to $4\sqrt{N}$.
To avoid frustration effects in the vortex lattice phase we use a
hexagonal simulation cell with periodic boundary conditions in all
directions.  In \Fig{fig.2} we show the helicity modulus in the
direction of the applied field $\Upsilon_z$, and the structure
function $S_\bq = \left<\left|m_z(\bq)\right|^2\right>/(NL_z)^2$ at a
reciprocal vector $\bq$ of the Abrikosov vortex lattice, as a function
of temperature.  $\Upsilon_x=\Upsilon_y=0$ for all $T$, reflecting the
absence of vortex pinning.  At the transition both $\Upsilon_z$ and
$S_\bq$ drop quite sharply, suggesting a first order melting
transition to an entangled vortex liquid with no intermediate
disentangled phase.  Right at the transition, the time series of the
internal energy or the structure function, obtained from the
simulation, fluctuate around two different values, giving further
support for a first order transition.  The inset shows how the average
energy per vortex and layer approaches a jump at the transition as
system size increases, with a latent heat of roughly $0.0015NL_zK$.
Taking into account the internal temperature dependence of the
parameters~\cite{Dogson}, and using values for YBCO gives an entropy
jump $\Delta S \approx 0.5\kB$ per vortex and layer, in rough
agreement with experiments~\cite{experiment}.

We now comment on the implications of the ideas presented above on the
analogy between the statistical mechanics of vortex systems and zero
temperature quantum field theory of bosons in
$(2+1)D$~\cite{Nelson,Fisher-Lee}.  The vortex lines are the analogue
of boson world lines in a path integral representation of the
partition function, with the $z$-direction playing the role of
imaginary time.  A crystal ground state for the bosons corresponds to
the Abrikosov vortex lattice phase in the vortex problem, while a
superfluid boson ground state is mapped to an entangled vortex line
liquid.  The winding number fluctuations of world lines in $(2+1)D$
gives the superfluid density of the boson problem~\cite{Ceperley}.  It
is of interest to study the consequences of our new area term $H'$ in
this context.  To this end it is useful~\cite{Feigelman} to view the
London interaction, as being mediated by a (massive in the screened
case) $(2+1)D$ gauge field $a_\mu$, see \Eq{2+1}.  The area term $H'$
is now generated by integrating over the ${\bf k=0}$ component of the
dual field strength $\tilde{f}_\mu=\epsilon_{\mu\nu\rho}\partial_\nu
a_\rho$ (which corresponds to the auxiliary field $\bb$ above \Eq{2+1}
in the vortex problem).  By coupling the dual field strength,
$\tilde{f}_\mu = (e_y, e_x, b_z)$, to an external source,
$\tilde{g}_\mu=(d_y,d_x,h_z)$, we find that the dual magnetic
permeability $\mu=\partial\left<b_z\right>/\partial h_z$ corresponds
to the helicity modulus in $z$ direction, $\Upsilon_z$
\cite{Blatter,Chen-Teitel}.  The inverse dual dielectric constants
$\epsilon_i^{-1}=\partial\left<e_i\right>/\partial d_i$ in the $x$ and
$y$ directions, correspond to the helicity modulus $\Upsilon_\mu$ in
the $y$ and $x$ directions, respectively.

In summary, we have shown how to include ${\bf k=0}$ fluctuations in
vortex line models, and how the helicity modulus can be obtained from
fluctuations of the projected area of vortex loops. This is useful for
detecting superconducting phase transitions in models with and without
screening of the London interaction.  We also presented continuum
Monte Carlo simulation results with a full 3D London interaction.
This model should be appropriate for the vortex lattice melting in
moderately anisotropic systems such as YBCO.  Similar approaches
should be useful in studies of, e.g., vortex glass transitions and
quantum phase transitions.

We thank Steve Girvin and Peter Olsson for valuable discussions.  This
work was supported by the Swedish Natural Science Research Council,
and by the Swedish Council for Planning and Coordination of Research
(FRN) and Parallelldatorcentrum (PDC), Royal Institute of Technology.


\begin{references}
\bibitem{foot} The form of the area term after summing over the global
moves may be related to the work in Ref.~\cite{Bokil-Young}.  Note
that global moves should not be included if dynamical properties of
the system are being considered.
\end{references}
\end{document}